\documentstyle[aps,prl,eqsecnum,multicol,epsfig]{revtex}

\renewcommand{\narrowtext}{\begin{multicols}{2} \global\columnwidth20.5pc}
\renewcommand{\widetext}{\end{multicols} \global\columnwidth42.5pc}
\newcommand{\Lrule}{\vspace*{-0.2in}\noindent\vrule width3.5in height.2pt
  depth.2pt \vrule depth0em height1em}
\newcommand{\Rrule}{\vspace{-0.1in}\hfill\vrule depth1em height0pt \vrule
  width3.5in height.2pt depth.2pt\vspace*{-0.1in}}

\begin{document}
\draft
\title{Topological universality of level dynamics in quasi-one-dimensional \\
disordered conductors}
\author{E. Kanzieper$^{1}$ and V. E. Kravtsov$^{1,2}$}
\address{$^{1}$ The Abdus Salam International Centre for Theoretical Physics, P.O. Box 586, 34100 Trieste, Italy \\
$^{2}$ Landau Institute for Theoretical Physics, 2 Kosygina str., 117940 Moscow, Russia}
\date{2 July 1999}
\maketitle

\begin{abstract}
Nonperturbative, in inverse Thouless conductance $g^{-1}$, corrections to distributions
of level velocities and level curvatures in quasi-one-dimensional disordered
conductors with a topology of a ring subject to a constant vector potential are studied 
within the framework of the instanton approximation of nonlinear $\sigma$-model. It is 
demonstrated that a global character of the perturbation reveals the universal features 
of the level dynamics. The universality shows up in the form of weak topological
oscillations of the magnitude $ \sim e^{-g}$ covering the main bodies of the 
densities of level velocities and level curvatures. The period of discovered universal
oscillations does not depend on microscopic parameters of conductor, and is only determined
by the global symmetries of the Hamiltonian before and after the perturbation was applied. 
We predict the period of topological oscillations to be $4/\pi^2$ for the distribution 
function of level curvatures in orthogonal symmetry class, and $\sqrt{3}/\pi$ for the 
distribution of level velocities in unitary and symplectic symmetry classes. 
\end{abstract}


\narrowtext

\section{Introduction}

Parametric level statistics describes the response of the spectrum $\{E_n\}$ of
complex chaotic systems to an external perturbation, being a measure of the sensitivity
of the individual energy level to a change of some external parameter $x$. The 
part of the parameter $x$ can be played by either the external electric or magnetic field, 
or the `strength' of a background potential. It may be any other relevant parameter  
of the Hamiltonian. 

In the series of papers \cite{SzA-1993,SA-1993,SLA-1993}, it was shown that a system whose 
spectrum follows closely the universal fluctuations predicted by the Wigner--Dyson random 
matrix theory \cite{M-1991} (RMT) should also exhibit a universal parametric behavior. 
This conclusion was reached by analyzing 
the correlations of the single electron level densities in a disordered metallic 
sample with the topology of a ring pierced by an Aharonov--Bohm magnetic flux $\varphi$. 
In this particular problem, the parametric correlations take a universal form 
involving the rescaled parameter $X^2 \propto g \varphi ^2$, with the dimensionless
conductance $g=E_c/\Delta$, the ratio of the Thouless energy and the mean level spacing. 
Numerical simulations \cite{SA-1993} have supported the point that the universal character of 
parametric level statistics extends to a wider class of chaotic systems without disorder
(chaotic billiards) whose Hamiltonian depends on some external parameter $x$. In such 
systems, the spectral fluctuations taken at different values of $x$ become 
system-independent after the rescaling, $x \mapsto X$, which involves solely the 
`generalized' dimensionless conductance $g \propto \Delta^{-2} \left\langle 
[ \partial E_n(x)/\partial x ]_{x=0}^2 \right\rangle$. [Here, the angular brackets
$\left\langle ... \right\rangle$ stand for disorder averaging.] Further investigations 
\cite{O-1994,FS-1995,THSA-1994} have established the universality of other parametric 
statistics such as the distribution functions $P_v(V)=\Delta \langle \sum_n \delta (V-V_n)
\delta (E-E_n(0)) \rangle$ and $P_c(K)=\Delta \langle \sum_n \delta (K-K_n) \delta 
(E-E_n(0)) \rangle$ of the level 
velocities $V_n = \Delta^{-1} [\partial E_n (x)/\partial x]_{x=0}$ and the level 
curvatures $K_n = \Delta^{-1} [\partial ^2 E_n (x)/\partial x^2]_{x=0}$. These two measures
of level dynamics turn out to be universal {\it after} the rescaling 
$V={\tilde V} \langle V_n^2\rangle ^{1/2}$ and $K={\tilde K} \langle |K_n|\rangle$. 
Both distributions \cite{Rem1}
\begin{mathletters}
\begin{eqnarray}
P_v(V) \propto \left\{ 
\begin{array}{cc}
\delta ({\tilde V}) , & \beta =1, \\ 
\exp \{ -{\tilde V}^2/2\} , & \beta =2,4,
\end{array}
\right.  
\label{vel-rmt}
\end{eqnarray}
and
\begin{eqnarray}
P_c(K) \propto (1+{\tilde K}^2) ^{-1-\beta /2}
\label{curv-rmt}
\end{eqnarray}
\end{mathletters}
appear to depend only on the Dyson index \cite{M-1991} $\beta=1,2$ and $4$. In turn, the averages
$\langle V_n^2 \rangle$ and $\langle |K_n| \rangle$ are deeply connected, via the 
Edwards--Thouless relation \cite{ET-1972}, to the average
system conductance if the parameter $x$ represents the Aharonov--Bohm flux. The latter 
circumstance makes it possible to use these rather abstract measures of level dynamics 
for distinguishing between the extended and the localized phases in disordered systems.

The above universality (that emerges after appropriate rescaling) implies that the 
resulting parametric correlations do not depend on 
details of the system and the perturbation. Instead, they are only determined by the 
fundamental (orthogonal, unitary or symplectic) symmetries of the system before and 
after it was perturbed. For disordered conductors, this universal regime is observed 
in the limit \cite{E-1983} of the infinite dimensionless conductance, $g = E_c/\Delta
\rightarrow \infty$. In quantum chaotic systems, the same criterion \cite{AASA-1996} 
applies provided one identifies the Thouless energy with the first nonzero 
mode $\gamma_1$ in the spectrum of the Perron--Frobenius operator of the corresponding 
classical system. It should be stressed that the extended, structureless electron 
eigenfunctions which do not correlate with the fluctuations of the electron energy 
levels are the underlying physical reasons of the universality phenomenon.

One-dimensional conductors in the strongly localized regime constitute another
extreme situation, in which the universality of parametric statistics is broken
completely. Recent studies \cite{CBSK-1996,TBF-1997} have shown that, in the regime of 
strong localization, the distribution function of level 
curvatures has nothing in common with the universal law Eq. (\ref{curv-rmt}). Instead,
it roughly follows the log-normal law which explicitly contains such system-dependent
parameters as the circumference $L$ of a one-dimensional ring and the electron mean free 
path $l$ (which is of order of the localization length $\xi$ for a one-channel
conductor).

Disordered conductors with large but finite values of $g$, which are in between of the
two extreme situations described above, offer a fertile field for tracing the mechanism 
of breaking the parametric universality. While, in this case, the electron wave functions 
are still extended, they do have a pronounced internal structure (showing up in their 
long-range correlations \cite{BM-1997,PA-1998}) that cannot already be neglected. 
Finite--$g$ 
corrections to parametric level statistics beyond the Wigner--Dyson RMT have been a focus 
of a number of studies \cite{AA-1995,YK-1997,BCKY-1998}. All of them have 
emphasized a {\it nonuniversal} character of the level dynamics manifesting itself
already in the form of the perturbative corrections \cite{YK-1997} to the level curvature distribution 
which becomes sensitive to both the spatial dimensionality $d$ of the disordered conductor and 
the origin of the external perturbation. The latter determines the sign of the correction
which depends on the presence or absence of the global gauge invariance (that is, on the
topological origin of the perturbation) and is different for the $T$--breaking 
perturbations in the form of random magnetic fluxes that act locally and for
the $T$--breaking perturbations represented by a constant vector potential (that is 
equivalent to a global phase twist in the boundary conditions). Nonanalyticity \cite{KY-1997}
of the level curvature distribution in the latter case in the low-dimensional
conductors ($d=1,2$) was also addressed.

In the present paper we further examine the level dynamics in quasi-one-dimensional disordered
conductors in the presence of the Aharonov--Bohm magnetic flux to demonstrate that, for
finite conductance $g$, a new kind of {\it topological universality} emerges, for which
the existence of the global gauge invariance is crucial. This is a nonperturbative in
$g^{-1}$ effect. To be precise, we argue that 
the ring topology and the global character of the perturbation are responsible for 
appearance of the weak {\it oscillations} covering the main bodies, Eqs. (\ref{vel-rmt}) and
(\ref{curv-rmt}), of the distributions of the level velocities and the level 
curvatures. While the magnitude of the oscillations is system dependent
(being of order of $\sim e^{-g}$), their period appears to be independent of 
microscopic parameters of the system and the perturbation: 
It is entirely determined by the global symmetries -- orthogonal, unitary or symplectic 
-- of the system before and after the perturbation was applied, and is different for level 
velocities and level curvatures. We also stress that, contrary to the Wigner--Dyson 
universality [Eqs. (\ref{vel-rmt}) and (\ref{curv-rmt})], that arises {\it after} 
rescaling $V \mapsto {\tilde V}$ and $K \mapsto {\tilde K}$, no rescaling is needed to 
establish the universality of topological oscillations: Their period is {\it universal} and 
parameter independent in {\it genuine} variables $V$ and $K$. In the absence of the
global gauge invariance, the above universality does not show up at all.

To appreciate the difference between the local and the global perturbations on the 
formal level, it is 
instructive to appeal to the random matrix formulation of the problem, replacing
the microscopic Hamiltonian by banded random matrices which are known \cite{FM-1991} 
to describe (in a certain thermodynamic limit) the physics of 
disordered quasi-one-dimensional conductors. For definiteness,
the orthogonal symmetry is supposed to be respected in the unperturbed conductor, while
the perturbation is assumed to drive the system symmetry toward the unitary one.

(i) For the {\it local}
perturbation of the strength $x$, the perturbed quasi-one-dimensional disordered system with $N$
sites can be modeled by the $N \times N$ parametric random matrix of the Pandey--Mehta
type \cite{PM-1983}
\begin{equation}
\label{h-local}
H^{\rm local}(x) = H^{R} + i x H^{A},
\end{equation}
where $H^{R}$ and $H^{A}$ are $N\times N$ real symmetric and real antisymmetric, 
statistically independent banded random matrices, respectively. Notice that such a 
decomposition imposes no constraints for the real
eigenvalues of the matrix $H^{\rm local}(x)$ as functions of the variable $x$. 

(ii) This contrasts the case of the {\it global}
perturbation applied to quasi-one-dimensional conductors with a ring topology which
should explicitly be incorporated \cite{CGIMZ-1994} to both unperturbed and perturbed matrix Hamiltonians. 
An unperturbed (real symmetric) matrix $H(0)$ must account for a periodic electron
motion along a discrete ring of $N$ sites (see Fig. 1).

\begin{figure}[-b]
\centerline{\epsfig{figure=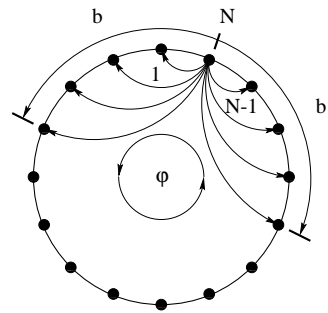,width=10pc}} 
\vspace*{5mm}
\caption{
A discrete ring of $N$ sites pierced by a (dimensionless) Aharonov--Bohm flux
$\varphi$. Connected geometry of a ring permits the electron hopping from the
$N$-th site to $b$ adjacent sites to the left and $b$ sites to the right.
}
\label{fig1}
\end{figure}

\begin{figure}[-b]
\centerline{\epsfig{figure=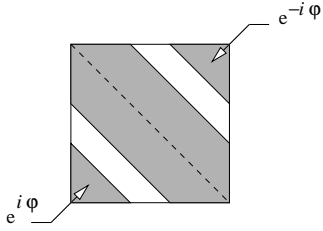,width=10pc}} 
\vspace*{5mm}
\caption{
Schematic representation of the structure of the periodic banded matrix that
contains three (shaded) regions with nonvanishing entries: The band of the width
$2b$ along the main diagonal, the upper right and the lower left corners (both
of the width $b$). In the presence of the Aharonov--Bohm flux, the corner entries
are multiplied by the phase factors $e^{\pm i\varphi}$, Eq. (\ref{h-global}).
}
\label{fig2}
\end{figure}

If the particle located at some site may 
reach only $b$ neighbors to the right and $b$ neighbors to the left ($b < [N/2]$), it
is easy to see that the unperturbed `periodic' $N \times N$ matrix $H(0)$ will contain 
three regions with nonvanishing entries: a band $|\mu-\nu| \le b$ along the main 
diagonal, and the upper right and the lower left corners, $|\mu-\nu| \ge N-b$ (see Fig. 2).

Let us now apply a small static magnetic flux $\Phi = (\varphi/2\pi) \Phi_0$, $\Phi_0 = h c/e$
being the flux quantum. Each time the electron hops from the 
site $\mu$ to the site $\mu+1$, it ascribes the phase $\varphi/N$. As a result, 
the transition amplitude between the sites $\mu$ and $\nu$ gets modified by the phase 
factor $e^{i\theta_{\mu\nu}}$, $H_{\mu\nu} \mapsto H_{\mu\nu} e^{i\theta_{\mu\nu}}$, with 
\begin{equation}
\label{phase-factor}
\theta_{\mu\nu}=\left\{ 
\begin{array}{cc}
[(\nu-\mu)/N]\varphi & \text{if } |\mu - \nu| \le b, \\ 
{\rm sgn} (\mu-\nu) [ 1 - |\nu-\mu|/N ]\varphi & \text{if } |\mu - \nu| \ge N - b.
\end{array}
\right.  
\end{equation}
Passing on to the basis, in which the eigenvectors of the perturbed matrix 
are transformed as $\psi_k \mapsto \psi_k e^{i (k/N)\varphi}$, we arrive at the perturbed 
periodic matrix of the form 
\begin{equation}
\label{h-global}
H_{\mu\nu}^{\rm global}(\varphi)=\left\{ 
\begin{array}{cc}
H_{\mu\nu}(0) & \text{if }|\mu-\nu| \le b, \\ 
H_{\mu\nu}(0)e^{-i\varphi} & \text{if }\nu-\mu \ge N - b, \\ 
H_{\mu\nu}(0)e^{i\varphi} & \text{if }\mu-\nu \ge N - b.
\end{array}
\right.  
\end{equation}
The perturbed periodic matrix $H^{\rm global}$ obtained in this way possesses
a manifestly different structure as compared to the structure of $H^{\rm local}$,
hereby reflecting the presence of the global gauge invariance in the former case.

The structure of Eq. (\ref{h-global}) dictates the $2\pi$--periodicity of the 
eigenvalues $E_n(\varphi)$ of $H^{\rm global}(\varphi)$, 
\begin{eqnarray}
\label{2pi}
E_n (\varphi+2\pi)=E_n(\varphi).
\end{eqnarray}
Equation (\ref{2pi}) suggests the Fourier analysis to be an appropriate 
language for description of level dynamics in the case of the global perturbation:
\begin{eqnarray}
\label{fourier}
E_n(\varphi) = \Delta\sum_{m=0}^{\infty} a_m^{(n)} \cos(m\varphi).
\end{eqnarray}
As a 
matter of fact, Eq. (\ref{fourier}) allows one to relate the problem of evaluating the 
Fourier transform $\tilde{P}_c(\lambda)=\int dK \exp\{-iK\lambda\} P_c(K)$ of the 
curvature distribution function to an effective statistical mechanical problem. To see 
this, we notice that the level curvature 
$K_n$ can be expressed in terms of the amplitudes $a_m^{(n)}$ as
\begin{equation}
\label{curv-ampl}
K_n = \frac{1}{\Delta}\left. \frac{\partial ^2 E_n}{\partial \varphi^2} 
\right|_{\varphi=0} = - \sum_{m=0}^{\infty} m^2 a_m^{(n)},
\end{equation}
so that the Fourier transform $\tilde{P}_c(\lambda)$ of the curvature distribution 
function is
\begin{equation}
\label{curv-fourier}
\tilde{P}_c(\lambda) = \Delta \left\langle
\sum_n \exp\left\{i\lambda \sum_{m=0}^{\infty} m^2 a_m^{(n)}\right\} \delta(E-E_n(0))
\right\rangle.
\end{equation}
Introducing the joint probability distribution function for the amplitudes $a_m^{(n)}$
in the form
\begin{equation}
\label{jpdf-ampl}
{\cal P}_g(\{a_m\}) = \Delta \left\langle
\sum_n \delta(E-E_n(0)) \prod_{m=0}^{\infty} \delta(a_m - a_m^{(n)})
\right\rangle,
\end{equation}
we conclude that the free energy $\tilde{F}_c(\lambda) = - \ln \tilde{P}_c(\lambda)$
of the effective statistical mechanical model equals
\begin{eqnarray}
\label{free-energy}
\tilde{F}_c(\lambda) &=& - \lim_{M \rightarrow \infty} \ln \int \prod_{m=0}^M da_m
{\cal P}_g(\{ a_m \}) 
\nonumber \\
&\times& \exp\left\{ i\lambda \sum_{m=0}^M m^2 a_m \right\}.
\end{eqnarray}
Here, ${\cal P}_g(\{a_m\})$ describes the interaction of fictitious `particles' $a_m$
subject to external `field' $i\lambda$. [Similar representation can be obtained for
the distribution function of the level velocities $V_n$ for $\beta=2,4$. In this
case, $\cos(m\varphi)$ is to be replaced by $\sin(m\varphi)$ in the Fourier expansion
Eq. (\ref{fourier}).] It is important to emphasize that no such mapping to an
effective statistical mechanical problem is available for the case of the local 
perturbation, Eq. (\ref{h-local}).

In the rest of the paper, we argue, by using the instanton approximation 
\cite{MK-1995,KY-1997} of nonlinear $\sigma$-model, that in the limit $g \gg 1$ the effective statistical mechanical 
model Eq. (\ref{free-energy}) exhibits a second-order phase transition as the external 
field $i\lambda$ varies, with the free energy $\tilde{F}_c(\lambda)$ displaying a 
discontinuity of its second derivative at the `critical' point determined by the value of 
the critical parameter $\lambda_c = \pi^3/2$. It turns out, that the value of the critical 
field $i\lambda_c$ is universal: It does not depend on microscopic
parameters of the problem but is entirely determined by the global character of the 
perturbation, Eq. (\ref{h-global}), that breaks the orthogonal symmetry toward the 
unitary one. This finding, being translated to the space of genuine curvature 
variable $K$, implies appearance of the universal oscillations with the 
{\it universal} period $\delta K = 2\pi/\lambda_c = 4/\pi^2$ and the system-dependent
amplitude
$\sim e^{-g}$ on the main body Eq. (\ref{curv-rmt}) of the level curvature distribution.
Similar analysis of the distribution 
functions of the level velocities (at $\beta =2$ and $4$) reveals the existence of the 
second-order phase transition in associated effective statistical mechanical problem 
as well. In the latter problem, the value of the critical external field $i\lambda_c$ is 
also found to be universal though different: $\lambda_c = 2\pi^2/\sqrt{3}$. As a result,
the main body Eq. (\ref{vel-rmt}) of the distribution function of the level velocities
gets dressed by the weak oscillations of the same magnitude ($\sim e^{-g}$) with the 
{\it universal} period $\delta V = 2\pi/\lambda_c = \sqrt{3}/\pi$. We stress that the
above universality has a clear topological origin: A global character of the perturbation
(constant vector potential) and a ring geometry of a quasi-one-dimensional conductor are 
necessary conditions for its existence.

The paper is organized as follows. Section II is devoted to a nonperturbative analysis 
of the level curvature distribution function associated with the global $T$--breaking 
perturbation over the orthogonal ensemble. The distribution of the level velocities in
ensemble with the unitary symmetry is analyzed in Section III. In Section IV a relationship
between the ring geometry of a conductor, the boundary conditions in associated nonlinear 
$\sigma$-model (in its quasi-one-dimensional and ergodic limits) and the discovered 
topological oscillations is discussed. Section V contains conclusions. The
details concerning the distribution function of the level velocities in the case of 
symplectic symmetry are collected in Appendix.

\section{Distribution of level curvatures at $\beta = 1$}

Let us consider the electron of mass $m$ moving in a disordered conductor with a topology
of a quasi-one-dimensional ring of the circumference $L$. In the presence of the magnetic flux $\Phi$ that pierces the ring, the one-particle
Hamiltonian takes the form
\begin{eqnarray}
\label{hamiltonian}
{\hat H} = \frac{\hbar^2}{2m}\left( i {\bf \nabla}_{\bf r} + \frac{\varphi}{L} {\bf e}_{\theta}
\right)^2 + V({\bf r}),
\end{eqnarray}
where $\varphi = 2\pi \Phi/\Phi_0$, $\Phi_0 = hc/e$ is the flux quantum, and
${\bf e}_{\theta}$ is the unit vector along the azimuthal direction. The potential
$V({\bf r})$ describes the effect of disorder and is supposed to be the $\delta$-correlated
random Gaussian process with $\langle V({\bf r})V({\bf r'})\rangle = (\hbar/2\pi \nu \tau)
\delta({\bf r} - {\bf r'})$. Here, $\tau$ is the mean time between collisions and $\nu$ 
is the density of states. The (dimensionless) magnetic flux $\varphi$ in the
Hamiltonian breaks the time-reversal invariance which has been respected in the absence 
of the perturbation.

In what follows we are interested in the distribution function 
$P_c(K) = \Delta \langle \sum_n \delta(K-K_n) \delta(E-E_n(0))\rangle$
of the level curvatures
\begin{eqnarray}
\label{curv}
K_n = \frac{2}{\Delta} \lim_{\varphi \rightarrow 0} \frac{E_n(\varphi)-E_n(0)}{\varphi^2}
\end{eqnarray}
fluctuating, for a given level, over the ensemble of disorder realizations. These fluctuations
are due to fluctuations of both the single electron eigenfunctions and the electron 
eigenlevels in an unperturbed system. [Here, $\Delta$ is the mean level spacing.] Distribution function of
level curvatures can be expressed
through the parametric two-level correlation function $R(\omega,\varphi)=\Delta^2 \langle
\nu(E+\omega,\varphi) \nu(E,0)\rangle$ by means of the limiting procedure 
\cite{YK-1997}
\begin{eqnarray}
\label{limit}
P_c(K) = \lim_{\varphi \rightarrow 0} \frac{\varphi^2}{2} R\left( \omega = 
\frac{K\varphi^2 \Delta}{2}, \varphi
\right),
\end{eqnarray}
similar to the one suggested in Ref. \cite{KZ-1992} for the distribution of the level 
velocities. The relation Eq. (\ref{limit}) is based on the fact that the level
velocity vanishes identically in ensembles with the weakly broken orthogonal symmetry,
Eq. (\ref{vel-rmt}).

The parametric two-level correlation function $R(\omega,\varphi)$, and hence the
distribution function $P_c(K)$ of the level curvatures, can be represented 
in the form of the functional integral over the $8 \times 8$ supermatrix field $Q$
by using the Efetov's nonlinear $\sigma$-model \cite{E-1983}. The latter statistics
is given by \cite{FS-1995,YK-1997}
\begin{equation}
\label{dflc-susy}
P_c(K) = \lim_{\varphi \rightarrow 0} \varphi^2 {\rm Re} \int_{Q^2(x)=\openone_8} 
{\cal D}Q p_c[Q]
\exp\{-F[Q;K,\varphi]\}.
\end{equation}
Here, $p_c[Q]$ is some preexponent irrelevant for our further consideration that holds
with the exponential accuracy. The free energy functional $F[Q;K,\varphi]$ reads:
\begin{eqnarray}
\label{fe-susy}
F[Q;K,\varphi] &=& \frac{\pi g}{8} \int dx {\rm Str} \left( \partial_x Q - i
\varphi \left[ {\hat \tau}, Q\right] \right)^2 \nonumber \\
&+& \frac{i\pi K \varphi^2}{8}
\int dx {\rm Str} \left( \Lambda Q \right).
\end{eqnarray}
In the above formulas, $x = r/L\in (-1/2,+1/2)$ is the dimensionless coordinate measured 
along the ring, $g = \hbar D/L^2\Delta$ is the dimensionless conductance, $g \gg 1$. 

For the ring topology the supermatrix field $Q(x)$ must obey the periodic boundary
conditions. Its symmetry properties are specified in Ref. \cite{E-1983}, so that $Q(x)$
spans the three subspaces: retarded--advanced (R-A), boson--fermion (B-F), and the subspace 
related to the time reversal (T-R). The supertrace, ${\rm Str}$, is defined as 
${\rm Str} M = {\rm Tr}M_{FF} - {\rm Tr}M_{BB}$. Other matrices entering 
Eq. (\ref{fe-susy}) are
\begin{eqnarray}
\label{tau-lambda}
{\hat \tau} = \frac{\openone_8 + \Lambda}{2}{\hat \tau_3},
\end{eqnarray}
with
\begin{eqnarray}
\label{aux-matrices}
{\hat \tau_3} &=& {\rm diag} \left( \sigma_z, \sigma_z, \sigma_z, \sigma_z  \right)_{\rm R-A},
\nonumber \\
\Lambda &=& {\rm diag} \left( \openone_2, \openone_2, -\openone_2, -\openone_2 \right)_{\rm R-A},
\end{eqnarray}
and 
\begin{eqnarray}
\label{sigma-z}
\sigma_z= \left( 
\begin{array}{cc}
1 & 0 \\ 
0 & -1
\end{array}
\right)
\end{eqnarray}
being the $z$-Pauli matrix in the T-R space. The matrix $\Lambda$ describes breaking the symmetry 
between the retarded and the advanced subspaces, while the matrix ${\hat \tau_3}$ breaks the
time reversal symmetry.

Notice that the operator $\partial_x - i\varphi [{\hat \tau},...]$ in Eq. (\ref{fe-susy})
implies the 
presence of the global gauge invariance due to the {\it global} character of the 
applied $T$--breaking perturbation represented by the Aharonov--Bohm flux $\varphi$. This 
operator gets destroyed if the perturbation has a {\it local} origin, that is the case
of random magnetic fluxes or magnetic impurities. In this local case, the term
linear in $\varphi$ is absent in Eq. (\ref{fe-susy}).

It was realized in Refs. \cite{KY-1997} that the level 
curvature $K$ enters the free
energy functional Eq. (\ref{fe-susy}) in the same way as does $\omega$ in the problem of 
the long-time current relaxation \cite{MK-1995}. This observation suggests that the 
instanton approximation of nonlinear $\sigma$-model should be applicable to 
the evaluation of the Fourier transform, Eq. (\ref{curv-fourier}), of the level curvature 
distribution:
\begin{eqnarray}
\label{cf}
{\tilde P}_c(\lambda) &=& \lim_{\varphi \rightarrow 0} \varphi^2 {\rm Re} \nonumber \\
&\times& 
\int dK \int_{Q^2(x)
=\openone_8} {\cal D}Q
p_c[Q] e^{-\left(F[Q;K,\varphi] +iK\lambda \right)} \nonumber \\
&\propto& {\rm Re} \; e^{-(F[Q_{\rm ins};K_{\rm sp},0] + i K_{\rm sp}\lambda)}.
\end{eqnarray}
Equation (\ref{cf}) holds with the exponential accuracy, and is valid whenever the
functional $F[Q_{\rm ins};K_{\rm sp},0] + i K_{\rm sp}\lambda$ is large enough. 
[Hereafter, we consider only positive values of $\lambda$ because of the evenness
of $P_c(K)$.]

An important remark is appropriate here. In accordance with the results of 
Ref. \cite{MK-1995} one can expect that at large enough $\lambda$ the space-independent 
configurations of $Q(x)$ are energetically unfavorable: There are essentially 
space-dependent, {\it periodic} superfields $Q_{\rm ins}(x)$ which minimize the 
functional in question. In contrast, at relatively small $\lambda$ we expect that 
the minimum of $F[Q;K,\varphi] + i K\lambda$ in the limit $\varphi\rightarrow 0$
is reached on spatially independent configurations of $Q(x)$. 
This formal argumentation suggests 
the existence of some `critical' point $\lambda = \lambda_c$ at which a transition 
between the two solutions occurs. Below, this 
scenario will be confirmed by explicit calculations. It turns out that
at the critical point the functional $F[Q_{\rm ins};K_{\rm sp},0]+iK_{\rm sp}\lambda$
and its first derivative with respect to $\lambda$ are continuous functions of $\lambda$,
and only the second derivative exhibits a discontinuity with immediate implications for 
the distribution function $P_c(K)$ itself to be discussed later on.

Two observations are to be made before we proceed with the computation of 
${\tilde P}_c(\lambda)$. First, we note that 
due to the limit $\varphi \rightarrow 0$ to be taken in Eq. (\ref{cf}) the only 
nonzero contribution to ${\tilde P}_c(\lambda)$ can
arise from the entries in the noncompact boson--boson sector of the supermatrix
$Q(x)$ because they are allowed to take arbitrary large values. Second, the 
integration over the Grassmann entries of $Q(x)$ modifies only the preexponent $p_c[Q]$
in Eq. (\ref{cf}) and, therefore, they do not contribute to 
${\tilde P}_c(\lambda) = \exp\{-{\tilde F}_c(\lambda)\}$ if the latter is evaluated 
with the exponential accuracy, provided ${\tilde F}_c(\lambda) \gg 1$. For these two 
reasons, we may retain only the commuting entries in the noncompact sector of
$Q(x)$ in Eq. (\ref{cf}). Hence, the supermatrix $Q$ may be thought of as a
conventional $4 \times 4$ matrix $Q_B(x)$.

Using the Efetov's parameterization \cite{E-1983} for the orthogonal symmetry class, we 
obtain
\begin{equation}
\label{o-param}
Q_B\left( x\right) =
U(x)
\left( 
\begin{array}{cc}
\cos {\hat \theta}_B & i\sin 
{\hat \theta }_B \\ -i\sin {\hat \theta }_B & -\cos {\hat \theta }_B
\end{array}
\right)_{{\rm R-A}} 
U^{-1}(x),
\end{equation}
where
\begin{eqnarray}
\label{u-matrix}
U(x)= \left( 
\begin{array}{cc}
e^{i\phi \sigma _z} & 0 \\ 
0 & e^{i\chi \sigma _z}
\end{array}
\right)_{\rm R-A}
\end{eqnarray}
and
\begin{mathletters}
\label{cos-sin}
\begin{eqnarray}
\label{theta-b}
\cos {\hat \theta}_B &=&
\cosh \theta_1 \cosh \theta_2 \openone_2 + \sinh \theta_1 \sinh\theta_2 \sigma_x, 
\label{cosine} \\
\sin {\hat \theta}_B &=& 
i\sinh \theta_1  \cosh \theta_2 \openone_2 + i\cosh \theta_1 \sinh\theta_2 \sigma_x.
\label{sine}
\end{eqnarray}
\end{mathletters}
Here, 
\begin{eqnarray}
\label{sigma-x}
\sigma_x = \left( 
\begin{array}{cc}
0 & 1 \\ 
1 & 0
\end{array}
\right)
\end{eqnarray}
is the $x$-Pauli matrix in the T-R space; 
$z$-Pauli matrix was specified in Eq. (\ref{sigma-z}).

In the equations above, $\phi, \chi \in (0,2\pi)$ and 
$\theta_1,\theta_2 \in (0,+\infty)$ are spatially dependent, 
{\it periodic} functions of the variable $x \in (-1/2,+1/2)$. Obviously, this
periodicity reflects the ring topology of the disordered conductor in
question. Substituting $Q_B(x)$ in
the parameterization Eqs. (\ref{o-param}) -- (\ref{cos-sin}) into Eq. (\ref{fe-susy}),
and varying the functional $F[Q;K,\varphi] + iK\lambda$ with respect to the functions
$\theta_1(x)$, $\theta_2(x)$, $\phi(x)$ and $\chi(x)$ we find a set of four 
saddle-point equations which are satisfied by the choice
$\theta_1(x) = \theta_2(x) = \theta(x)$
and $\phi(x) = {\rm const}$. The functions $\theta(x)$ and $\chi(x)$ obey the
following saddle-point equations:
\begin{equation}
\label{theta-eq}
\frac{d^2}{dx^2}[2\theta(x)] + \sinh[2\theta(x)]\left[ \kappa \varphi^2 -
\left( \frac{d}{dx}\chi(x) - \varphi \right)^2 \right]=0, 
\end{equation}
\begin{equation}
\label{hi-eq}
\frac{d}{dx}\left[ \left( \frac{d}{dx}\chi(x) -\varphi \right)
\sinh^2\theta(x)\right]=0,
\end{equation}
where we have introduced $\kappa = iK_{\rm sp}/2g$. Also, variation over $K$ yields the 
first self-consistency equation
\begin{equation}
\label{self}
\varphi^2 \int dx \cosh^2 \theta(x) = 2\lambda.
\end{equation}
In terms of solutions to the above saddle-point equations, the functional ${\tilde F}_c(\lambda)
= F[Q_{\rm ins};K_{\rm sp},0] + i K_{\rm sp}\lambda$ reads
\widetext
\Lrule
\begin{eqnarray}
\label{fe-1}
{\tilde F}_c(\lambda) = \lim_{\varphi \rightarrow 0} \left\{ \pi g \int dx
\left[
\left( \frac{d\theta}{dx} \right)^2 + \sinh^2 \theta(x) \left( \frac{d\chi}{dx} -
\varphi\right)^2 -\kappa \varphi^2 \cosh^2 \theta(x)
\right] + 2 g \kappa \lambda \right\}.
\end{eqnarray}
\Rrule
\narrowtext
\noindent
Notice that Eq. (\ref{fe-1}) is justified as long as ${\tilde F}_c(\lambda) \gg 1$.

In order to perform the limit $\varphi \rightarrow 0$, we introduce two new functions
$v(x) = \chi(x)/\varphi$ and ${\tilde \theta}(x) = \theta(x) + \ln \varphi$ to reduce
Eqs. (\ref{theta-eq}) and (\ref{hi-eq}) to
\begin{eqnarray}
\label{theta-eq2}
\frac{d^2 {\tilde \theta}}{dx^2} + \frac{1}{4}e^{2{\tilde \theta}(x)}\left[
\kappa - \left(\frac{dv}{dx}-1 \right)^2
\right] = 0, \\
\label{v-eq}
\frac{d}{dx} \left[ \left( \frac{dv}{dx}-1\right) e^{2{\tilde \theta}(x)}
\right] = 0,
\end{eqnarray}
with ${\tilde \theta}(x) \in (-\infty,+\infty)$. The first self-consistency equation 
Eq. (\ref{self}) takes the form
\begin{eqnarray}
\label{self-2}
\int dx e^{2{\tilde \theta}(x)} = 8\lambda,
\end{eqnarray}
and
\begin{eqnarray}
\label{fe-2}
{\tilde F}_c(\lambda) = \pi g \int dx \left( d {\tilde \theta}/dx \right)^2 + 
2g\kappa \lambda.
\end{eqnarray}
When deriving Eq. (\ref{fe-2}), we have taken into account Eq. (\ref{theta-eq2}), as 
well as the periodicity of ${\tilde \theta}^{\prime}(x)$. 

Saddle-point equations Eqs. (\ref{theta-eq2}) and (\ref{v-eq}) correspond to the
{\it global} $T$--breaking perturbation as is set by Eqs. (\ref{hamiltonian}) and 
(\ref{fe-susy}). For the {\it local} perturbation, the linear in $\varphi$ term disappears
from the functional Eq. (\ref{fe-susy}) leading to the replacement of the terms
$(dv/dx-1)^2$ and $dv/dx-1$ by the terms $(dv/dx)^2 +1$ and $dv/dx$ in 
Eqs. (\ref{theta-eq2}) and (\ref{v-eq}), respectively. It can be shown \cite{BCKY-1998}
that the modified Eq. (\ref{v-eq}) and the periodicity of $v(x)$ result in the condition 
$dv/dx=0$ which, in turn, leaves the only possibility to satisfy the modified
Eq. (\ref{theta-eq2}) by a spatially homogeneous configuration of ${\tilde \theta}(x)$
at $\kappa=1$. This corresponds to the universal Wigner--Dyson regime (see also 
the discussion (i) after Eq. (\ref{pot})).

Now we are back to Eqs. (\ref{theta-eq2}) -- (\ref{fe-2}) which can be brought to a more 
convenient form. Resolving Eq. (\ref{v-eq}), and making use of the periodicity of $v(x)$ 
we obtain
\begin{eqnarray}
\label{sce2}
\int dx e^{-2{\tilde \theta}(x)} = {\cal N}.
\end{eqnarray}
This is the second self-consistency equation. Further, introducing the new notations
\begin{mathletters}
\label{nn}
\begin{eqnarray}
w(x) &=& {\tilde \theta}(x) - \xi, \label{nn-1} \\
e^{-\xi} &=& {\cal N} \sqrt{\kappa},
\label{nn-2}
\end{eqnarray}
\end{mathletters}
we rewrite the only remained saddle-point equation Eq. (\ref{theta-eq2}) as 
\begin{mathletters}
\label{spe}
\begin{eqnarray}
\frac{d^2 w}{dx^2} + \sigma^2 \sinh w(x) = 0, \label{meq} \\
\sigma^2 = \sqrt{\kappa}/{\cal N}. \label{sig}
\end{eqnarray}
\end{mathletters}
It should be supplemented by the two self-consistency equations
which we combine to
\begin{eqnarray}
\label{sces}
\int dx e^{\pm w(x)} = \frac{1}{\sqrt{\kappa}}
= \frac{8\lambda {\cal N} \sqrt{\kappa}}{\pi}.
\end{eqnarray}
Here, the periodicity of $w^{\prime}(x)$ has explicitly been taken into account.
Also,
\begin{eqnarray}
\label{fef}
{\tilde F}_c(\lambda) = \frac{\pi g}{4}\int dx \left( \frac{dw}{dx}\right)^2
+ 2 g \kappa \lambda.
\end{eqnarray}

Equation (\ref{spe}) is an equation of motion of a classical 
particle with a unit mass in the potential
\begin{eqnarray}
U(w) = \sigma^2[\cosh w -1],
\label{pot} 
\end{eqnarray}
$w$ and $x$ playing the role of coordinate and time, respectively. Due to a ring 
topology of our problem, we are 
interested in those solutions to Eq. (\ref{spe}) that describe the finite motion of
the particle with the period $T=1$. There are two kinds of solutions available. 

(i) The first solution is a trivial one, $w(x) = 0$. In this case, one concludes from 
Eq. (\ref{sces}) that the saddle-point value of $K_{\rm sp}$ is given by $\kappa =1$. Equation
(\ref{fef}) then tells us that ${\tilde F}_c(\lambda) \equiv {\tilde F}_c^{{\rm WD}}(\lambda)= 2g\lambda$ for arbitrary
$\lambda \ge 0$. One can 
verify, by making use \cite{Rem2} of Eq. (\ref{curv-rmt}) of Introduction for $\beta=1$, that this 
trivial solution appears to describe (with the exponential accuracy) the distribution of 
the level curvatures in the universal, Wigner--Dyson regime. 

(ii) There is, however, the second solution which is
spatially inhomogeneous. The important point here is that the finite motion in the 
potential $U(w)$ cannot occur with an arbitrary period. It is bounded from above, 
$T({\cal E}) \le T_{\rm max} = 2\pi/\sigma$, with
\begin{equation}
\label{per-mot}
T({\cal E}) = \sqrt{2} \int_{w_{-}({\cal E})}^{w_{+}({\cal E})} \frac{dw}{\sqrt{{\cal E}-U(w)}}
= \frac{4{\bf K}(k)}{\sigma}(1-k^2)^{1/4}.
\end{equation}
Here, $w_{\pm}$ are the turning points of the finite motion, $U(w_{\pm}) = {\cal E}$,
and ${\cal E}$ is the integral of motion
\begin{eqnarray}
\label{iom}
{\cal E} = \frac{1}{2} \left( \frac{dw}{dx} \right)^2 + \sigma^2 [\cosh w -1].
\end{eqnarray}
The parameter $k \in (0,1)$ is defined by
\begin{eqnarray}
\label{k-dep}
k^2 = \frac{2\sqrt{\left( 1+ {\cal E}/\sigma^2\right)^2-1}}
{\left( 1+ {\cal E}/\sigma^2\right) + \sqrt{ \left(1+ {\cal E}/\sigma^2
\right)^2 - 1}},
\end{eqnarray}
${\bf K}(k) = {\bf F}(\pi/2,k)$, with 
\begin{equation}
\label{elliptic}
{\bf F}(\psi,k) = \int_{0}^{\psi} \frac{d\alpha}{\sqrt{1-k^2\sin^2\alpha}}
\end{equation}
being the elliptic integral of the first kind \cite{RG}.

The maximal period 
$T_{\rm max}$ corresponds to the almost harmonic particle oscillations in a
vicinity of the minimum $w=0$ of the potential $U(w)$. Alternatively, for a given
period $T=1$ (which we are interested in), the nontrivial period--$1$ solution
exists only for $\sigma \le \sigma_c = 2\pi$. It is given by
\begin{mathletters}
\label{sol-2}
\begin{eqnarray}
x - \frac{1}{2} &=& - \frac{{\bf F}(\psi(x),k)}{2{\bf K}(k)}, \label{s1} \\
\psi(x) &=& \arcsin\left( \frac{1}{k} \sqrt{1-\sqrt{1-k^2}e^{w(x)} } \;\right).
\label{s2}
\end{eqnarray} 
\end{mathletters}

Equation (\ref{sol-2}) makes it possible to complete our computational program. To
this end, we substitute Eq. (\ref{sol-2}) into Eq. (\ref{sces}) to obtain
\begin{eqnarray}
\label{sce-w}
\int dx e^{\pm w(x)} = \frac{{\bf E}(k)}{\sqrt{1-k^2} {\bf K}(k)}.
\end{eqnarray}
Here, ${\bf E}(k)= {\bf E}(\pi/2,k)$ is the elliptic integral of the 
second kind \cite{RG}
\begin{eqnarray}
\label{ell-2}
{\bf E}(\psi,k) = \int_{0}^{\psi} d\alpha \sqrt{1-k^2 \sin^2 \alpha}. 
\end{eqnarray}
Further, we insert Eq. (\ref{sol-2}) into Eq. (\ref{fef}) and take
into account Eqs. (\ref{iom}), (\ref{sce-w}), (\ref{sces}), 
(\ref{sig}) as well as the identity
\begin{eqnarray}
\sigma = 4{\bf K}(k)(1-k^2)^{1/4}
\label{iden}
\end{eqnarray}
following from Eq. (\ref{per-mot}) taken at $T=1$, to derive after some algebra
the instanton solution ${\tilde F}_c(\lambda) \equiv {\tilde F}_c^{\rm ins}(\lambda)$ in 
the form
\begin{eqnarray}
\label{action}
{\tilde F}_c^{{\rm ins}}(\lambda) = 4\pi g {\bf K}(k) \left[
(2-k^2){\bf K}(k) -{\bf E}(k)
\right].
\end{eqnarray}  
\begin{figure}[-b]
\centerline{\epsfig{figure=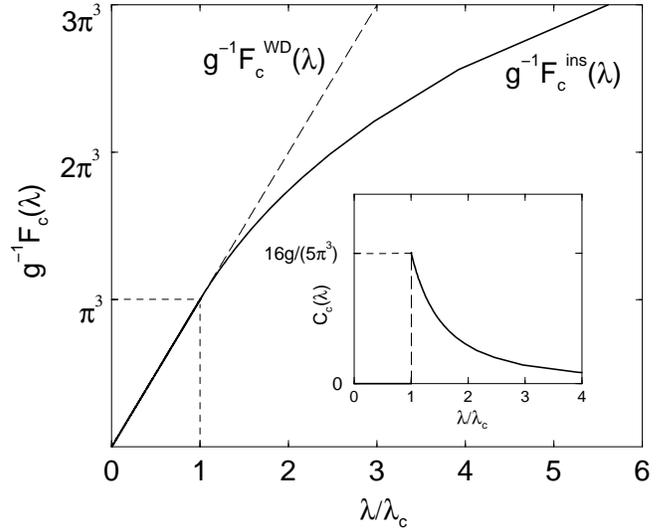,width=20pc}} 
\vspace*{5mm}
\caption{
The function ${\tilde F}_c(\lambda)$ given by Eq. (\ref{fs}). The dashed line
depicts the Wigner--Dyson solution extended to the region $\lambda > \lambda_c$.
A jump of the `heat capacity' $C_c(\lambda)$ at the critical point $\lambda=\lambda_c$ 
is shown in the inset.
}
\label{fig3}
\end{figure}
It remains to find the relationship between the initial parameter $\lambda$ of
the Fourier space in the l.h.s. of Eq. (\ref{action}) and the auxiliary parameter
$k$. This is done by combining Eqs. (\ref{sig}), (\ref{sces}), (\ref{sce-w}) and
(\ref{iden}) with the result
\begin{eqnarray}
\label{lambda-k}
\lambda(k) = \frac{2\pi}{1-k^2} \frac{{\bf E}^3(k)}{{\bf K}(k)}.
\end{eqnarray}

Let us analyze Eqs. (\ref{action}) and (\ref{lambda-k}) which are manifestly different
from ${\tilde F}_c^{{\rm WD}}(\lambda) = 2g\lambda$ with $\lambda \ge 0$. It was already
argued that the nontrivial solution Eq. (\ref{sol-2}) occurs at $\sigma \le \sigma_c =
2\pi$. As is seen from Eqs. (\ref{iden}) and (\ref{lambda-k}), this domain maps to 
$0 \le k \le 1$ with $k_c = 0$, and/or to $\lambda \ge \lambda(0)=\lambda_c$, with
\begin{eqnarray}
\label{lambda-c}
\lambda_c = \frac{\pi^3}{2}.
\end{eqnarray}
Exactly at the point $\lambda=\lambda_c$ one observes that 
${\tilde F}_c^{{\rm ins}}(\lambda_c)
\equiv {\tilde F}_c^{{\rm WD}}(\lambda_c)$, while at $\lambda \ge \lambda_c$ the
inequality ${\tilde F}_c^{{\rm ins}}(\lambda) \le {\tilde F}_c^{{\rm WD}}(\lambda)$ 
holds. [In particular, it follows from Eqs. (\ref{action}) and (\ref{lambda-k}) that
\cite{KY-1997} ${\tilde F}_c^{\rm ins}(\lambda) = \pi g \ln^2 \lambda$ as $\lambda \gg 1$.] This
means that for $\lambda \ge \lambda_c$ it is the spatially inhomogeneous, periodic 
solution Eq. (\ref{sol-2}) to the saddle-point equation
Eq. (\ref{spe}) that minimizes the functional 
$F[Q;K,\varphi] + i K\lambda$ in Eq. (\ref{cf}) as $\varphi\rightarrow 0$.
Therefore, we are forced to conclude that
\begin{eqnarray}
\label{fs}
{\tilde F}_c(\lambda) = \left\{ 
\begin{array}{cc}
2 g \lambda & \text{for }0\le \lambda < \lambda_c, \\ 
{\tilde F}_c^{\rm ins}(\lambda)
 & \text{for }\lambda
\ge \lambda _c.
\end{array}
\right.  
\end{eqnarray}
The result Eq. (\ref{fs}) supports the qualitative arguments given after Eq. (\ref{cf}).

A distinctive feature of Eq. (\ref{fs}), which holds as long as 
$g \gg 1$, is that the second derivative of
${\tilde F}_c(\lambda)$ with respect to $\lambda$ exhibits a
discontinuity exactly at the critical point $\lambda=\lambda_c$,
in whose vicinity $|\lambda-\lambda_c| \ll \lambda_c$ we derive
\begin{eqnarray}
\label{expansion}
\frac{{\tilde F}_c(\lambda)}{g} = 2\lambda - \frac{2\pi^3}{5}
\Theta(\lambda-\lambda_c)\left( 1- \frac{\lambda}{\lambda_c}\right)^2.
\end{eqnarray}
One can define the `heat capacity' $C_c(\lambda) = -d^2{\tilde F}_c/d\lambda^2$
which undergoes a jump at $\lambda =\lambda_c$: $C_c(\lambda)=0$ for $0 \le \lambda < \lambda_c$,
and $C_c(\lambda) = 16 g /5\pi^3$ for $\lambda = \lambda_c + 0$. This is the same 
discontinuity which is characteristic of the usual second-order
phase transitions treated in the mean-field approximation, the factor $g$
playing the role of the `volume'. Equation (\ref{expansion}) is an evidence to the
second-order phase transition in the effective statistical mechanical model 
Eq. (\ref{free-energy}) proposed in the Introduction.

The above discontinuity (revealed in the Fourier space $\lambda$) has an 
important impact upon the distribution of the level curvatures. Being
translated to the space of genuine curvature variable $K$, Eq. (\ref{expansion}) 
implies appearance
of the weak oscillations covering the main body of the distribution
function of the level curvatures, with a system-dependent amplitude of order 
$\sim e^{-g}$ and the
{\it universal} period
\begin{eqnarray}
\label{curv-period}
\delta K = \frac{2\pi}{\lambda_c} = \frac{4}{\pi^2}.
\end{eqnarray}
We stress that the value of $\delta K$ does not depend on microscopic parameters of 
the Hamiltonian Eq. (\ref{hamiltonian}). Its value is only determined by the ring
topology of the disordered conductor as well as by the 
orthogonal symmetry of the problem weakly broken by the global external perturbation
in the form of a constant vector potential.

We close this section by noticing that the above universal oscillations do not
appear for the local perturbation. In that case, neither the statistical mechanical
model Eq. (\ref{free-energy}) can be introduced, nor are the nontrivial periodic
configurations of the $Q$-matrix available in associated nonlinear $\sigma$-model 
\cite{BCKY-1998}. This signals of a topological origin of the discovered oscillations.

\section{Distribution of level velocities at $\beta = 2$}

In this section, we extend the previous analysis to study manifestations of the
gauge invariance in the distribution function of the single electron level velocities 
in quasi-one-dimensional disordered ring where the time reversal symmetry has been broken from 
the very beginning so that the additionally applied small magnetic field respects
the initial (unitary) symmetry of the problem. As is the case of the curvature 
distribution, the distribution $P_v(V) = \Delta \langle \sum_{n}
\delta(V-V_n)\delta(E-E_n(0))\rangle$ of the level velocities 
\begin{eqnarray}
\label{vel-vel}
V_n = \frac{1}{\Delta} \lim_{\varphi \rightarrow 0} \frac{E_n(\varphi) -E_n(0)}{\varphi}
\end{eqnarray}
can be expressed through the parametric
two-level correlation function $R(\omega,\phi)$ by means of the limiting procedure
\begin{equation}
\label{vel-def}
P_v(V) = \lim_{\varphi \rightarrow 0} \varphi R\left( \omega =V\varphi \Delta,\varphi 
\right),
\end{equation}
as first proposed in Ref. \cite{KZ-1992}. 

We take the conventional route, and express the parametric two-level correlation function 
in terms of the nonlinear $\sigma$-model. The desired distribution of level velocities is
\begin{equation}
\label{lv-susy}
P_v(V) = \lim_{\varphi \rightarrow 0} \varphi {\rm Re} \int_{Q^2(x)=\openone_8}
{\cal D}Q p_v[Q] \exp\left\{ -F[Q;V,\varphi]\right\}.
\end{equation}
Here, $p_v[Q]$ is some preexponent which is irrelevant for our calculations performed
with the exponential accuracy. The free energy functional in Eq. (\ref{lv-susy}) reads
\begin{eqnarray}
\label{vel-fun}
F[Q;V,\varphi] &=& \frac{\pi g}{8} \int dx {\rm Str} \left(
\partial_x Q - i \varphi [{\hat \tau},Q]
\right)^2 \nonumber \\
&+& \frac{i \pi V \varphi}{4} \int dx {\rm Str} (\Lambda Q).
\end{eqnarray}
Despite the similarity of Eq. (\ref{vel-fun}) to Eq. (\ref{fe-susy}), the 
symmetry properties \cite{E-1983} of the $8\times 8$ supermatrix field $Q$ are different in these two 
situations. [In particular, under assumption that the time reversal symmetry is completely 
destroyed, one has \cite{SA-1993,AIE-1993} $[Q,{\hat \tau_3}] = 0$.]

Evaluation of the Fourier transform of the distribution function 
of level velocities is the main goal of this section. In the instanton approximation 
used above one obtains (with the exponential accuracy)
\begin{equation}
\label{vel-fourier}
{\tilde P}_v(\lambda) = \int dV P_v(V) e^{-iV\lambda} \propto {\rm Re} \;
e^{-(F[Q_{{\rm ins}}; V_{\rm sp},0] + i V_{\rm sp}\lambda)},  
\end{equation}
where $\lambda$ is considered to be positive, $\lambda \ge 0$, owing to the evenness of
$P_v(V)$. Let us proceed as follows. Due to the limit $\varphi \rightarrow 0$ to be 
implemented, we have to retain only noncompact degrees of freedom in the supermatrix
$Q$ which are given by its boson--boson sector. Simultaneously, we omit the Grassmann 
entries, as 
discussed in the previous section. As a result, we replace $8\times 8$ supermatrix $Q$
by conventional $4 \times 4$ matrix $Q_B$ in the
parameterization of Eqs. (\ref{o-param}) and (\ref{u-matrix}), in which
\begin{mathletters}
\label{theta-b-un}
\begin{eqnarray}
\cos {\hat \theta}_B &=&
\cosh \theta_1 \openone_2,\label{cosine-un} \\
\sin {\hat \theta}_B &=& 
i\sinh \theta_1 \openone_2.
\label{sine-un}
\end{eqnarray}
\end{mathletters}
Here, $\phi,\chi \in (0,2\pi)$ and $\theta_1 \in (0,+\infty)$ are spatially
dependent, {\it periodic} functions of the variable $x \in (-1/2,+1/2)$.
Substituting $Q_B(x)$ into Eq. (\ref{vel-fun}),
and varying the functional $F[Q;V,\varphi]+iV\lambda$ with respect to the 
functions $\theta_1(x)=\theta(x)$, $\phi(x)$ and $\chi(x)$, we find the following set 
of the saddle-point equations that determine the instanton configuration $Q_{\rm ins}$:
\begin{equation}
\label{sp-eq-un-1}
\frac{d^2\theta}{dx^2} + \sinh \theta(x) \left[
\frac{\kappa}{2} \varphi - \left(\frac{d}{dx}\chi(x) - \varphi \right)^2
\cosh\theta(x) 
\right] = 0, 
\end{equation}
\begin{eqnarray}
\label{sp-eq-un-2}
\frac{d}{dx}\left[ \left( \frac{d}{dx} \chi(x) - \varphi
\right)\sinh^2 \theta(x)\right]=0.
\end{eqnarray}
Here, we have introduced $\kappa = 2iV_{\rm sp}/g$. Also, one has $\phi(x)={\rm const}$.

Variation over $V$ yields the first self-consistency equation
\begin{eqnarray}
\label{sc-un-3}
\pi \varphi \int dx \cosh\theta(x) = \lambda.
\end{eqnarray}
In terms of the above functions, the functional ${\tilde F}_v(\lambda)
= F[Q_{\rm ins}; V_{\rm sp},0] + i V_{\rm sp}\lambda$ takes the form
\widetext
\Lrule
\begin{equation}
\label{un-functional}
{\tilde F}_v(\lambda) = \lim_{\varphi\rightarrow 0} \left\{
\frac{\pi g}{2} \int dx \left[ \left( \frac{d\theta}{dx}\right)^2 
+ \sinh^2\theta(x)\left( \frac{d\chi}{dx} - \varphi\right)^2
- \kappa \varphi \cosh\theta(x)
\right] + \frac{1}{2}g\kappa\lambda
\right\}.
\end{equation}
\Rrule
\narrowtext
\noindent
To implement the limit $\varphi \rightarrow 0$, we perform the following transformation,
$v(x)= \chi(x)/\varphi$ and ${\tilde \theta}(x) = \theta(x) + \ln \varphi$, that results
in two saddle-point equations
\begin{eqnarray}
\label{sp-u-1eq}
\frac{d^2 {\tilde \theta}}{dx^2}+\frac{1}{4}e^{{\tilde \theta}(x)}
\left[
\kappa - \left( \frac{dv}{dx}-1\right)^2 e^{{\tilde \theta}(x)}
\right]=0,
\end{eqnarray}
\begin{eqnarray}
\label{sp-u-2eq}
\frac{d}{dx}\left[
\left( \frac{dv}{dx}-1
\right)e^{2{\tilde \theta}(x)}
\right]=0,
\end{eqnarray}
with ${\tilde \theta}(x)\in (-\infty,+\infty)$. The first self-consistency equation is
\begin{eqnarray}
\label{sp-u-eq3}
\pi \int dx e^{{\tilde \theta}(x)} = 2\lambda,
\end{eqnarray}
while the functional ${\tilde F}_v(\lambda)$ equals
\begin{eqnarray}
\label{sp-u-eq4}
{\tilde F}_v(\lambda) = \frac{\pi g}{2}\int dx \left( \frac{d{\tilde \theta}}{dx}
\right)^2 + \frac{1}{4} g \kappa \lambda.
\end{eqnarray}
In deriving Eq. (\ref{sp-u-eq4}) we have used the periodicity of ${\tilde \theta}^{\prime}(x)$,
and Eqs. (\ref{un-functional}), (\ref{sp-u-1eq}) and (\ref{sp-u-eq3}). Second self-consistency
equation is obtained from Eq. (\ref{sp-u-2eq}),
\begin{eqnarray}
\label{sce-ecs}
\int dx e^{-2{\tilde \theta}(x)} = {\cal N}.
\end{eqnarray}

It is convenient to pass on to the new variables 
\begin{mathletters}
\label{w-not}
\begin{eqnarray}
w(x) &=& \frac{3}{2} {\tilde \theta}(x) + \xi, \label{w-not-1} \\
e^{\xi} &=& {\cal N} \sqrt{\kappa}, \label{w-not-2}
\end{eqnarray}
\end{mathletters}
to reduce the saddle-point equation Eq. (\ref{sp-u-1eq}) to
\begin{mathletters}
\label{w-eq}
\begin{eqnarray}
\frac{d^2w}{dx^2} + \sigma^2 e^{-w(x)/3} \sinh w(x) = 0, \label{w-eq-01} \\
\sigma^2 = \frac{3}{4}\left(\frac{\kappa}{{\cal N}} \right)^{2/3}. \label{w-eq-02}
\end{eqnarray}
\end{mathletters}
The two self-consistency equations take the form
\begin{eqnarray}
\label{sc-w-vel}
\int dx e^{-4w(x)/3} &=& \int dx e^{2w(x)/3} \nonumber \\
&=& \frac{1}{{\cal N}^{1/3} \kappa^{2/3}}
= \frac{2\lambda}{\pi} {\cal N}^{2/3}\kappa^{1/3}.
\end{eqnarray}
Also,
\begin{eqnarray}
\label{fv}
{\tilde F}_v(\lambda) = \frac{2}{9} \pi g \int dx \left( \frac{dw}{dx} \right)^2
+\frac{1}{4} g \kappa \lambda.
\end{eqnarray}

Equation (\ref{w-eq}) describes a motion of a classical particle of a unit mass
in the potential
\begin{equation}
\label{cm-vel}
U(w) = \frac{3\sigma^2}{8} \left[ 2e^{2w/3} + e^{-4 w/3} - 3\right],
\end{equation}
$w$ playing the part of the coordinate, and $x$ being the time. Due to the Aharonov--Bohm
geometry of the initial problem, we are interested in the periodic solutions to 
Eq. (\ref{w-eq}), $T=1$. 

(i) One such solution is a trivial one, $w(x) = 0$. In this case, we conclude from the 
two self-consistency equations Eq. (\ref{sc-w-vel}) that ${\cal N}^{1/3}\kappa^{2/3} = 1$ 
and $2\lambda {\cal N}^{2/3} \kappa^{1/3} = \pi$. It then follows that the saddle-point
value of $V_{\rm sp}$ is given by $\kappa = 2\lambda/\pi$. Substituting
the latter result into Eq. (\ref{fv}) we get 
${\tilde F}_v(\lambda) \equiv {\tilde F}_v^{\rm WD}(\lambda) = g\lambda^2 /2\pi$ for all $\lambda \ge 0$. One can 
verify, by using \cite{Rem3} Eq. (\ref{vel-rmt})
of Introduction, that the trivial solution $w=0$ corresponds to the universal Wigner--Dyson
form of the distribution of the level velocities at $\beta=2$. 

(ii) There is, however, a nontrivial period--$1$ solution to Eq. (\ref{w-eq}) 
provided $\lambda \ge \lambda_c$, $\lambda_c$ being some universal 
constant. Depending on the integral of motion
\begin{eqnarray}
\label{cofm}
{\cal E} = \frac{1}{2} \left( \frac{dw}{dx}\right)^2 + \frac{3\sigma^2}{8}
\left[
2e^{2w/3} + e^{-4w/3} - 3
\right],
\end{eqnarray}
the spatially inhomogeneous solution to Eq. (\ref{w-eq}) has the period
\begin{eqnarray}
\label{period-vel}
T({\cal E}) &=& \sqrt{2} \int_{w_{-}({\cal E})}^{w_{+}({\cal E})} \frac{dw}{\sqrt{{\cal E}
-U(w)}} \nonumber \\
&=& \frac{2\sqrt{6}}{\sigma \sqrt{\gamma_0-\gamma_{-}}}{\bf K}(k).
\end{eqnarray}
Here, $w_{\pm}({\cal E})$ are the turning points, $U(w_{\pm})={\cal E}$. Notations
in Eq. (\ref{period-vel}) are as follows:
\begin{mathletters}
\label{notations}
\begin{eqnarray}
k^2 &=& \frac{\sqrt{3}-\tan(\alpha/3)}
{\sqrt{3}+ \tan(\alpha /3)}, \label{not-1} \\
\gamma_0 &=& \frac{1}{2\sin^{2/3}(\alpha/2)} \left[
1+2\cos\left( \frac{\alpha}{3}\right)\right], \label{ng0} \\
\gamma_{\pm} &=& \frac{1}{2\sin^{2/3}(\alpha/2)} \left[
1-2\cos\left( \frac{\alpha \pm \pi}{3} \right)\right],  
\label{ngpm} \\
\cos\alpha &=& 1- 2\left( 1+ \frac{8{\cal E}}{9\sigma^2} \right)^{-3} 
\label{not-2}.
\end{eqnarray}
\end{mathletters}
If we fix the period of a finite motion to be $T=1$, the period--$1$
solution (that exists only for $\sigma \le \sigma_c = 2\pi$) is given by
\begin{mathletters}
\label{sol-vel}
\begin{eqnarray}
x+ \frac{1}{2} &=& \frac{{\bf F}(\psi(x),k)}{2{\bf K}(k)}, \label{sv-1} \\
\psi(x) &=& \arcsin \left(\frac{1}{k} \sqrt{\frac{e^{2w(x)/3} - \gamma_{+}}
{e^{2w(x)/3} - \gamma_{-}}}\;
\right).
\label{sv-2}
\end{eqnarray}
\end{mathletters}

At this stage, we are able to find a closed analytical expression for 
${\tilde F}_v(\lambda)$ defined by Eq. (\ref{fv}). To this end, we substitute
the solution Eq. (\ref{sol-vel}) into Eq. (\ref{sc-w-vel}) to derive
\begin{equation}
\label{sc-www}
\int dx e^{2w(x)/3} = \gamma_{+} + (\gamma_{+} - \gamma_{-})\left[
\frac{{\bf E}(k)}{(1-k^2){\bf K}(k)} -1
\right].
\end{equation}
Further, we insert Eq. (\ref{sol-vel}) into Eq. (\ref{fv}), and take into account
Eqs. (\ref{cofm}), (\ref{sc-w-vel}), (\ref{sc-www}), (\ref{w-eq-02}) as well as the
identity
\begin{eqnarray}
\label{iden-un}
\sigma = \frac{2\sqrt{6}}{\sqrt{\gamma_0-\gamma_{-}}}{\bf K}(k)
\end{eqnarray}
following from Eq. (\ref{period-vel}) taken at $T=1$, to obtain after some
algebra the instanton solution ${\tilde F}_v(\lambda) \equiv {\tilde F}_v^{\rm ins}(\lambda)$ 
in the form
\widetext
\Lrule
\begin{eqnarray}
\label{f-velocity}
{\tilde F}_v^{{\rm ins}}(\lambda) = \frac{8\pi g}{\gamma_{0}-\gamma_{-}}{\bf K}^2(k)
\left[
\gamma_{-}+\gamma_0
- 
(\gamma_{+}-\gamma_{-})\left(
\frac{{\bf E}(k)}{(1-k^2){\bf K}(k)} - 1
\right)
\right].
\end{eqnarray}
Here, the parameter $\lambda$ in the l.h.s. of Eq. (\ref{f-velocity}) is connected to 
the auxiliary parameter $k \in (0,1)$ as
\begin{eqnarray}
\label{lambda-vel}
\lambda (k) = \frac{2\pi \sqrt{2}}{\sqrt{\gamma_0 - \gamma_{-}}}{\bf K}(k)
\left[
\gamma_{+} + (\gamma_{+}-\gamma_{-})\left(\frac{{\bf E}(k)}{(1-k^2){\bf K}(k)}-1
\right)
\right]^2.
\end{eqnarray}
\Rrule
\narrowtext
\noindent
The above relationship is obtained by combining Eqs. (\ref{w-eq-02}), (\ref{sc-w-vel}),
(\ref{sc-www}) and (\ref{iden-un}).

Let us analyze the solution derived. It obviously differs from ${\tilde F}_v^{{\rm WD}}
= g\lambda^2/2\pi$ with $\lambda \ge 0$ (that is related to a trivial case $w(x)=0$). 
First, the instanton solution Eq. (\ref{f-velocity}) arises at $\sigma \le \sigma_c =2\pi$
that corresponds to $\lambda \ge \lambda(0) = \lambda_c$,
\begin{eqnarray}
\label{lambda-vel-c}
\lambda_c = \frac{2\pi^2}{\sqrt{3}}.
\end{eqnarray}
Second, exactly at the point $\lambda=\lambda_c$, one observes that ${\tilde F}_v^{\rm ins}
(\lambda_c) = {\tilde F}_v^{\rm WD}(\lambda_c)$, while at $\lambda \ge \lambda_c$ the
inequality ${\tilde F}_v^{\rm ins}
(\lambda) \le {\tilde F}_v^{\rm WD}(\lambda)$ is fulfilled. [In particular, one has
\cite{MK-1995}
${\tilde F}_v(\lambda)=2\pi g \ln^2 \lambda$ as $\lambda \gg 1$.] 
This means that for $\lambda
\ge \lambda_c$ the spatially inhomogeneous, periodic solution Eq. (\ref{sol-vel}) to
the saddle-point equation Eq. (\ref{w-eq}) leads to such configurations $Q_{\rm ins}$
which become energetically more favourable as compared to the trivial solution $w=0$. 
As a result, we conclude that
\begin{eqnarray}
\label{final-vel-u}
{\tilde F}_v(\lambda) = \left\{ 
\begin{array}{cc}
g \lambda^2/2\pi & \text{for }0\le \lambda < \lambda_c, \\ 
{\tilde F}_v^{\rm ins}(\lambda) & \text{for }\lambda
\ge \lambda _c.
\end{array}
\right.  
\end{eqnarray}
The most important feature of Eq. (\ref{final-vel-u}), which is valid for $g \gg 1$, is 
that the second
derivative of ${\tilde F}_v(\lambda)$ with respect to $\lambda$ exhibits a
discontinuity exactly at the critical point $\lambda=\lambda_c$, in whose vicinity
$|\lambda-\lambda_c|\ll \lambda_c$ the following expansion holds
\begin{eqnarray}
\label{exp-vel-u}
\frac{{\tilde F}_v(\lambda)}{g} = \frac{\lambda^2}{2\pi} - \frac{12\pi^3}{11}
\Theta(\lambda-\lambda_c) \left(1-\frac{\lambda}{\lambda_c} \right)^2.
\end{eqnarray}
The `heat capacity' $C_v(\lambda) = -d^2 {\tilde F}_v/d\lambda^2$ is seen to posses 
a jump at $\lambda=\lambda_c$: $C_v(\lambda_c +0)-C_v(\lambda_c - 0)= 18g/11\pi$.

The above discontinuity, similar to the one observed in usual second-order phase
transitions, signals that the main body of the distribution function of the level 
velocities gets dressed by weak oscillations with the amplitude of
order $\sim e^{-g}$. Their period is {\it universal},
\begin{eqnarray}
\label{period-v-un}
\delta V = \frac{2\pi}{\lambda_c} = \frac{\sqrt{3}}{\pi},
\end{eqnarray}
for it does not depend on microscopic parameters of the disordered conductor. The 
value $\delta V$ is only determined by the unitary symmetry of the system
and by the global type of the external perturbation applied.

\section{Ring topology: Quasi-one-dimensional versus 
Wigner--Dyson limit}

The analysis presented indicates that the universally oscillating corrections 
are of a topological origin: A ring geometry of a conductor subject to a constant 
vector potential (that represents a global perturbation and sets a global gauge invariance
on the level of the microscopic description) is a necessary condition for their existence.
However, the very fact of appearance of oscillations with a {\it universal} period, raises
the question about {\it actual} importance of quasi-one-dimensionality for the above
effect. 

Indeed, if we suppose that the ring topology (literally taken in its 
{\it geometric} sense, Fig. 1) is the only condition required to observe the topological 
universality, it could be a plausible assumption that, as long as $g \gg 1$, the same
oscillating features must persist in the distribution functions of level curvatures and 
level velocities in the ergodic regime. In the language of the random matrix theory, this 
is equivalent to the assumption that invariant random matrices closed in a ring (see below)
should lead to the same results, provided the dimensionless conductance $g$ is replaced by
the effective conductance $g(N)$ linearly scaling with the matrix dimension $N$,
$g(N) \sim N$. Below we argue that this is by no means
the case. It turns out that quasi-one-dimensionality is an essential ingredient of the
theory, being responsible for the forming the periodic boundary conditions in associated 
nonlinear $\sigma$-model. It is these boundary conditions which ultimately led to 
identifying the universal critical point $\lambda_c$ in the Fourier transforms of 
the distribution functions in question and motivated our claim about existence of the
second-order phase transition in the statistical mechanical model 
Eq. (\ref{free-energy}).

To support this point, it is useful to develop a discrete $\sigma$-model approach 
to the periodic banded matrices defined by Eq. (\ref{h-global}). The advantage of this
class of random matrices lies in the fact that, in the thermodynamic limit, they offer a 
possibility of probing the two physically different regimes depending on the behavior of 
the scaling parameter $\eta=b^2/N$ as $N \rightarrow \infty$. 
Namely, (i) If the band width $b$ scales with the matrix dimension $N$ as 
$b = g_{*}^{1/2}N^{1/2}$, $g_{*} \sim {\cal O}(N^0)$, the banded random matrices 
correspond to the quasi-one-dimensional physics ($\eta \rightarrow g_{*}$); 
(ii) If the band width $b$ scales linearly with $N$, $b=\alpha N$, $0 < \alpha \le 1$, 
the banded random matrices flow toward the ergodic, Wigner--Dyson regime 
($\eta \propto \alpha^2 N \rightarrow \infty$). Below, only the main points of the derivation of
nonlinear $\sigma$-model are sketched. For simplicity, the unitary symmetry is assumed;
also, we set $\varphi = 0$.

Let us represent the ensemble of periodic random matrices Eq. (\ref{h-global}) 
in the form 
\begin{equation}
\label{pm-def}
H_{\mu\nu} = H_{\mu\nu}^{(1)} + H_{\mu\nu}^{(2)},
\end{equation}
where $H^{(1)}$ and $H^{(2)}$ are statistically independent, Hermitian, Gaussian
distributed matrices with the variances 
\begin{eqnarray}
\label{pm-variances}
\langle H_{\mu\nu}^{(1,2)} H_{\mu^{\prime}\nu^{\prime}}^{(1,2)}\rangle = 
\delta_{\mu\nu^{\prime}}
\delta_{\nu\mu^{\prime}} J_{\mu\nu}^{(1,2)}.
\end{eqnarray}
Here, $J^{(1)}_{\mu\nu}$ describes the correlations between the matrix elements belonging to 
the band of the width $2b$ along the main diagonal, and $J^{(2)}_{\mu\nu}$ is due to the
correlations between the elements of the right upper and the left lower corners in 
Eq. (\ref{h-global}) (see Fig. 2). To make further consideration as clear as possible, we specify
these correlation matrices as
\begin{mathletters}
\label{j-choice}
\begin{eqnarray}
J_{\mu\nu}^{(1)} &=& e^{-|\mu-\nu|/b}, \label{j1} \\
J_{\mu\nu}^{(2)} &=& 2 e^{-N/b} \cosh[(\mu-\nu)/b]. \label{j2}
\end{eqnarray}
\end{mathletters}
The choice Eq. (\ref{j-choice}) enables us to treat
the model of periodic banded matrices exactly. [In the case of conventional
banded matrices, where only $J_{\mu\nu}^{(1)}$ appears, this fact has been used in 
Ref. \cite{FM-1991}.]

To derive a nonlinear $\sigma$-model, we follow the standard route \cite{E-1983,VWZ-1985}.
The generating functional $Z({\cal J})$ reads
\begin{equation}
\label{gf}
Z({\cal J}) = \int d[\Psi] 
\left\langle e^{- i \overline{\Psi} \left( H^{(1)} + H^{(2)}\right)\Psi } \right\rangle
e^{i\overline{\Psi}\left( E+{\cal J}+\Lambda(\omega + i0)/2 \right)\Psi}.
\end{equation}
Here, ${\cal J}$ is the source term, $\Psi$ is $4N$ component supervector, with 
$\overline{\Psi} = \Psi^{\dagger}L$; 
$L = {\rm diag} (+1,-1,+1,+1)_{{\rm R-A}}$ and $\Lambda ={\rm diag}(+1,+1,-1,-1)_{{\rm R-A}}$. 
Performing the 
averaging $\langle ... \rangle$ over the ensembles of random matrices $H^{(1)}$ and 
$H^{(2)}$, one obtains
\begin{equation}
\label{gf-a}
Z({\cal J}) = \int d[\Psi] 
e^{i\overline{\Psi}\left( E+{\cal J}+\Lambda(\omega + i0)/2 \right)\Psi}
e^{-\frac{1}{2} \sum_{\mu\nu} J_{\mu\nu}{\rm Str} (q_\mu q_\nu)},
\end{equation}
where $J_{\mu\nu} \equiv J_{\mu\nu}^{(1)} + J_{\mu\nu}^{(2)}$, and 
$q_\mu = \Psi_\mu \otimes \overline{\Psi}_\mu$ respects the symmetry 
$q_\mu = L q_\mu^{\dagger}L$. The next step consists of decoupling the interaction term 
in Eq. (\ref{gf-a}) by means of the Hubbard--Stratonovich transformation, followed by
integrating out the supervector $\Psi$. Omitting the details, we show the result:
\begin{eqnarray}
\label{gf-hst}
Z({\cal J}) &=& \int \prod_{\mu=1}^N {\cal D}{\hat \sigma}_\mu 
e^{-\frac{1}{2} \sum_{\mu,\nu=1}^N [J^{-1}]_{\mu\nu}
{\rm Str}({\hat \sigma}_\mu {\hat \sigma}_\nu)} \nonumber \\
&\times&
\prod_{\mu=1}^N {\rm Sdet}^{-1} \left(
E+{\cal J}+ \frac{\omega+i0}{2}\Lambda -{\hat \sigma}_\mu
\right).
\end{eqnarray}

The following observation is important here: Since the interaction
term in the exponent of Eq. (\ref{gf-a}) depends only on that part of supermatrices $q_\mu$ which can
be expanded in the eigenvectors $O_\sigma (\mu)$ of the correlation matrix $J_{\mu\nu}$,
the auxiliary supermatrices ${\hat \sigma}_\mu$ appearing in the course of the 
Hubbard--Stratonovich transformation and entering Eq. (\ref{gf-hst}) have
necessarily to belong to the same class of supermatrices specified by the expansion
\begin{equation}
\label{expansion-2}
{\hat \sigma}_\mu = \sum_\sigma {\hat c}_\sigma O_\sigma (\mu),
\end{equation}
${\hat c}_\sigma$ being the supermatrix coefficients.
Equation (\ref{expansion-2}) implies, in turn, that the supermatrices ${\hat \sigma}_\mu$
are {\it not} allowed to obey arbitrary boundary conditions. The boundary conditions
are uniquely {\it fixed} by the structure of the correlation matrix $J_{\mu\nu}$ (or, equivalently, by the structure
of the inverse matrix $[J^{-1}]_{\mu\nu}$). 

Although the generating functional in the form of Eq. (\ref{gf-hst}) is only an 
intermediate (though exact) result in deriving a nonlinear $\sigma$-model, the conclusion
made about the boundary conditions in a discrete $\sigma$-model is quite general: The 
constrained supermatrices $Q_\mu$, $Q_\mu ^2 =\openone_4$, that appear on the later stage 
of the saddle-point evaluation of the $N$-fold supermatrix integral in Eq. (\ref{gf-hst}), 
must also belong to the same class, Eq. (\ref{expansion-2}). Hence, the boundary 
conditions for constrained supermatrices $Q_\mu$ are dictated by the structure of 
$[J^{-1}]_{\mu\nu}$ as well.

From now on, we are interested in the inverse matrix $[J^{-1}]_{\mu\nu}$. It possesses a 
remarkably simple tridiagonal structure supplemented by (generically) nonvanishing entries
$(1,N)$ and $(N,1)$: 
\widetext
\Lrule
\begin{eqnarray}
\label{inverse}
[J^{-1}]_{\mu\nu} \equiv \frac{1}{1-e^{-2/b}}\left[
(1+e^{-2/b})\delta_{\mu ,\nu}(1-\delta_{\mu , 1})(1-\delta_{\mu , N})
+ \frac{(1+e^{-N/b})^2 - e^{-4/b}}{(1+e^{-N/b})^2 - e^{-2/b}}(\delta_{\mu,1}
\delta_{\nu,1} + \delta_{\mu,N}\delta_{\nu,N}) \right. \nonumber \\
\left.
- e^{-1/b}(\delta_{\mu,\nu+1} + \delta_{\nu,\mu+1})
-e^{-1/b}(1-e^{-2/b})\frac{1+e^{-N/b}}{(1+e^{-N/b})^2-e^{-2/b}}(\delta_{\mu,1}
\delta_{\nu,N} + \delta_{\nu,1}\delta_{\mu,N})
\right].
\end{eqnarray}
\Rrule
\narrowtext
\noindent
At finite $N$, the entries $(1,N)$ and $(N,1)$ are the hallmark of the ring geometry 
incorporated in the random matrix Eq. (\ref{h-global}) at $\varphi =  0$.
 
Let us analyze the eigenvectors $O_\sigma (\mu)$ of the inverse correlation matrix
Eq. (\ref{inverse}) for two different thermodynamic limits (i) and (ii) introduced 
prior to Eq. (\ref{pm-def}).

(i) In the quasi-one-dimensional limit, $b=g_*^{1/2} N^{1/2}$, the normalized eigenvectors 
of $[J^{-1}]_{\mu\nu}$ are those of the matrix
\begin{eqnarray}
\label{jm-q1d}
2\delta_{\mu ,\nu} - (\delta_{\mu , \nu+1} + \delta_{\nu , \mu +1})
- (\delta_{\mu ,1}\delta_{\nu ,N} + \delta_{\nu ,1}\delta_{\mu ,N}),
\end{eqnarray}
which turns out to be independent of $g_*$. The eigenvectors are found to be 
\begin{eqnarray}
\label{ev-q1d}
O_\sigma (\mu )=\frac 1{
\sqrt{N}}(-1)^\mu e^{i\mu \theta _\sigma },
\end{eqnarray}
where
\begin{eqnarray}
\label{el-q1d}
\theta _\sigma =\left\{ 
\begin{array}{cc}
(2\pi /N)\sigma  & \text{for }N\text{ even,} \\ (2\pi /N)\left( \sigma
-1/2\right)  & \text{for }N\text{ odd,}
\end{array}
\right. 
\end{eqnarray}
and $1 \le \sigma \le N$.

It is seen from Eqs. (\ref{expansion-2}), (\ref{ev-q1d}) and (\ref{el-q1d}) that
the supermatrices ${\hat \sigma}_\mu$ obey the {\it periodic} boundary conditions
\begin{eqnarray}
\label{pbc}
{\hat \sigma}_{\mu + N} = {\hat \sigma}_\mu
\end{eqnarray}
as do the supermatrices $Q_\mu$ of the corresponding nonlinear $\sigma$-model. Thus, we
conclude that in the thermodynamic limit $N\rightarrow \infty$ with the fixed parameter 
$b N^{-1/2}$, the {\it geometric} periodicity
built in the matrix Hamiltonian Eq. (\ref{h-global}) is {\it equivalent} to the
periodic boundary conditions imposed on $Q_\mu$ in nonlinear $\sigma$-model. [On the
formal level, this is due to the $(1,N)$ and $(N,1)$ entries in Eq. (\ref{jm-q1d})
that remain finite as $N\rightarrow \infty$ and couple the fields ${\hat \sigma}_1$ 
and ${\hat \sigma}_N$ in the exponent of Eq. (\ref{gf-hst}).] 

(ii) In the limit $b=\alpha N$, that corresponds to approaching the ergodic, Wigner--Dyson
regime, the normalized eigenvectors of $[J^{-1}]_{\mu\nu}$ are those of the matrix
(compare with Eq. (\ref{jm-q1d}))
\begin{eqnarray}
\label{jm-wd}
2\delta_{\mu ,\nu} - (\delta_{\mu , \nu +1} + \delta_{\nu ,\mu +1})
- (\delta_{\mu, 1}\delta_{\nu ,1} + \delta_{\mu ,N}\delta_{\nu , N}).
\end{eqnarray}
Notice that the parameter $\alpha$ has dropped from Eq. (\ref{jm-wd}), and the entries
$\delta_{\mu ,1}\delta_{\nu ,N}$ and
$\delta_{\nu ,1}\delta_{\mu ,N}$ have disappeared,
leading to destroying the coupling between ${\hat \sigma}_1$ and ${\hat \sigma}_N$ 
in the exponent of Eq. (\ref{gf-hst}). The 
eigenvectors of Eq. (\ref{jm-wd}) are 
\begin{eqnarray}
\label{ev-wd}
O_\sigma (\mu) = \sqrt{\frac{2}{(1+\delta_{\sigma ,0})N}} \cos[(\pi \sigma/N)(\mu -1/2)],
\end{eqnarray}
where $0 \le \sigma \le N-1$.

An immediate consequence of the solution Eq. (\ref{ev-wd}) is that the periodic 
boundary conditions for ${\hat \sigma}_\mu$ (and hence for $Q_\mu$) are {\it not}
dictated by the above set of the eigenvectors $O_\sigma (\mu)$. Contrary to the
quasi-one-dimensional situation, we deduce from Eqs. (\ref{expansion-2}) and (\ref{ev-wd})
the boundary conditions of the closed disordered sample which (in the continuum limit) 
read:
\begin{eqnarray}
\label{cbc}
\left. \frac{\partial}{\partial \mu}{\hat \sigma}_\mu \right|_{\mu = -1/2} =
\left. \frac{\partial}{\partial \mu} {\hat \sigma}_\mu \right|_{\mu = N-1/2} = 0.
\end{eqnarray}
Equation (\ref{cbc}) requires the current across the effective boundary of the 
sample to vanish. [Notice that Eq. (\ref{cbc}) is a close analog of the so-called
`trapping plane' boundary conditions arising in the radiative transfer theory 
\cite{C-1960}.] Hence, we infer that in the thermodynamic
limit $N\rightarrow\infty$ with the parameter $b N^{-1}$ being fixed the connected geometry
of the matrix Hamiltonian Eq. (\ref{h-global}) is generically {\it not} equivalent to the 
periodic boundary conditions in the associated (nonlinear) $\sigma$-model.

Equations (\ref{pbc}) and (\ref{cbc}) are the main outcome of this Section. They
show that in the thermodynamic limit $N \rightarrow \infty$ the closed geometry 
of the matrix Hamiltonian Eq. (\ref{h-global}) does not necessarily induce the periodicity 
of the supermatrix fields in the associated nonlinear $\sigma$-model. The criterion
to distinguish between the two types of the boundary conditions follows from 
Eq. (\ref{inverse}) that enables us to estimate the relative magnitude of the
coupling entries $\delta_{\mu , 1}\delta_{\nu , N}$ ($\delta_{\nu , 1}\delta_{\mu , N}$)
which equals $b e^{-N/b}$ provided $1 \ll b \ll N$. The
fulfilment of the inequality $b e^{-N/b} \ll 1$ leads to the periodic boundary conditions;
vice versa, the inequality $b e^{-N/b} \gg 1$ correponds to the boundary conditions
of the closed sample. It was observed, that while such a periodicity
is respected in the quasi-one-dimensional limit (i), it breaks down 
in the ergodic (Wigner--Dyson) limit (ii) being naturally replaced by the boundary 
conditions of the isolated disordered sample. This latter circumstance rules out
the very possibility to set up the problem of parametric level statistics with a
gauge-type perturbation in the ergodic regime. For this reason, we have to conclude 
that the quasi-one-dimensionality is of vital importance for 
existence of the universal topological oscillations in the distribution functions of the 
level velocities and the level curvatures.

\section{Conclusions}

The consideration presented in this paper has been concentrated on two particular 
measures of parametric level statistics in {\it quasi-one-dimensional} conductors with a 
ring topology in the presence of a constant vector potential. They are the distribution 
functions of the level curvatures ($\beta=1$) and the level velocities ($\beta =2$ and $4$). 
We have used the instanton formulation of the reduced nonlinear $\sigma$-model to evaluate
nonperturbative, in inverse Thouless conductance $g^{-1}$, corrections to the universal
random-matrix-theory distribution functions Eqs. (\ref{vel-rmt}) and (\ref{curv-rmt}).
Our main findings are rather startling, especially in the light of the previous studies
on the subject, which have created a common belief that small but finite value of $g^{-1}$
inevitably results in {\it nonuniversal} corrections beyond RMT. Contrary to these
expectations, we have demonstrated that in quasi-one-dimensional disordered conductors
with $g \gg 1$ there exists another type of corrections which, to a large extent, are
{\it universal}. The corrections are of topological origin and manifest themselves in the form of the weak oscillations on the
main bodies of the distribution functions of the level curvatures and the level velocities.
While the magnitude of the topological oscillations is system dependent ($\sim e^{-g}$),
their period appears to be independent of microscopic parameters of
conductor, and hence universal. It is entirely determined by the global symmetries of the Hamiltonian before
and after the perturbation was applied. We have predicted the period of the oscillations
to be $4/\pi^2$ for the distribution of level curvatures at $\beta=1$, and $\sqrt{3}/\pi$
for the distribution of level velocities at $\beta=2$ and $4$. It should be stressed
that no rescaling is needed to establish the universality of the topological oscillations:
Their period is universal and parameter independent in genuine curvature and velocity 
variables.

\section*{Acknowledgments}
Stimulating discussions with B. L. Altshuler are appreciated with thanks.

\appendix
\def\theequation{A.\arabic{equation}}

\section*{Distribution of level velocities at $\beta =4$}

The appendix aims to demonstrate that the conclusions drawn for the distribution
function of the level velocities at $\beta=2$ remain valid for $\beta=4$. The starting 
point of our analysis is Eqs. (\ref{vel-fourier}) and
({\ref{vel-fun}}), in which the parameterization of the supermatrix $Q(x)$ must correspond
to the symplectic symmetry \cite{E-1983}. Due to the limit $\varphi \rightarrow 0$ to
be taken, we keep only boson--boson sector in $Q$; also, the exponential accuracy of our calculations
allows one to eliminate the Grassmann entries there. We, therefore, arrive at Eq. (\ref{o-param}),
with ${\hat \theta}_B$ given by Eq. (\ref{theta-b-un}); the matrix $U(x)$ is to be
replaced by
\begin{eqnarray}
\label{u-matrix-sym}
U(x)= \left( 
\begin{array}{cc}
\openone_2 & 0 \\ 
0 & F(x)
\end{array}
\right),
\end{eqnarray}
with
\begin{eqnarray}
F(x) &=& \frac{1}{D(x)} \label{f-sym-1} \\
&\times& \left[
(D(x)-2)\openone_2 + 2 i m \sigma_z + 2 i m_1^{\prime} \sigma_x + 2 i m_1^{\prime\prime}
\sigma_y
\right], \nonumber \\
D(x) &=& 1 + m^2(x) + m_1^{\prime 2}(x)+ m_1^{\prime\prime 2}(x). \label{f-sym-2}
\end{eqnarray}
Here, 
\begin{eqnarray}
\label{sigma-y}
\sigma_y= \left( 
\begin{array}{cc}
0 & -i \\ 
i & 0
\end{array}
\right)
\end{eqnarray}
is the $y$-Pauli matrix ($\sigma_x$ and $\sigma_z$ were specified by 
Eqs. (\ref{sigma-x}) and (\ref{sigma-z})), and the functions
$m(x), m_1^{\prime}(x), m_1^{\prime\prime}(x) \in (-\infty,+\infty)$. In the above
parameterization, the free energy ${\tilde F}_v(\lambda,\varphi) = F[Q;V,\varphi]+i V\lambda$ is
\widetext
\def\theequation{A.\arabic{equation}}
\begin{eqnarray}
{\tilde F}_v(\lambda,\varphi) &=& iV\lambda+\frac{\pi g}{2} \int dx \left\{
\left( \frac{d\theta}{dx}\right)^2 + \frac{i\pi V \varphi}{4} \cosh\theta(x)
+ \varphi^2 \sinh^2\theta(x) \right\}
+ 2\pi g \int dx \frac{\sinh^2 \theta(x)}{D^2(x)} \nonumber \\
&\times&
\left\{
\left[
\left( \frac{dm}{dx}\right)^2 + \left( \frac{dm_1^{\prime}}{dx}\right)^2
+ \left( \frac{dm_1^{\prime\prime}}{dx}\right)^2  
+
\varphi \left(
(2-D(x))\frac{dm}{dx} + m \frac{dD}{dx} + 2 m_1^{\prime}\frac{dm_1^{\prime\prime}}{dx}
-2 m_1^{\prime\prime} \frac{dm_1^{\prime}}{dx} 
\right)
\right]
\right\}.
\label{fe-sym-par}
\end{eqnarray}
In order to simplify the formulas, we perform the limit $\varphi \rightarrow 0$ in
Eq. (\ref{fe-sym-par}) prior to taking the variational derivatives. This is done by
absorbing $\varphi$ into $\theta(x)$, $\theta(x)= {\tilde \theta}(x) - \ln\varphi$,
followed by introducing the new functions $h(x)=-2m(x)/\varphi$, 
$h_1^{\prime}(x)=-2m_1^{\prime}(x)/\varphi$ and $h_1^{\prime\prime}(x)=-2m_1^{\prime\prime}
(x)/\varphi$. The limit $\varphi \rightarrow 0$ results in
\begin{eqnarray}
{\tilde F}_v(\lambda,0) = \frac{\pi g}{2} \int dx \left\{
\left( \frac{d{\tilde \theta}}{dx}\right)^2 + \frac{1}{4} e^{2{\tilde \theta}(x)}
\left[
\left(\frac{dh}{dx}-1 \right)^2 + \left(\frac{dh_1^{\prime}}{dx} \right)^2
+\left( \frac{dh_1^{\prime\prime}}{dx}\right)^2
\right]
\right\}
+iV\left( \lambda - \frac{\pi}{2} \int dx e^{{\tilde \theta}(x)}
\right).
\label{limit-sym}
\end{eqnarray}
Evaluation of the instanton action is by now standard. Taking the functional derivative
of Eq. (\ref{limit-sym}) with respect to the functions entering ${\tilde F}_v(\lambda,0)$, 
we
obtain the following set of the saddle-point equations:
\begin{eqnarray}
\frac{d^2 {\tilde \theta}}{dx^2} &-& \frac{1}{4} e^{2{\tilde \theta}(x)} \left[
\left( \frac{dh}{dx}- 1\right)^2 + \left( \frac{dh_1^{\prime}}{dx}\right)^2
+ \left( \frac{dh_1^{\prime\prime}}{dx}\right)^2
\right] 
+ \frac{\kappa}{4}e^{{\tilde \theta}(x)} = 0,
\label{sp-sym-a} 
\end{eqnarray}
\Rrule
\narrowtext
\noindent
\begin{eqnarray}
\frac{d}{dx} \left[ \left( \frac{dh}{dx}- 1
\right)e^{2{\tilde \theta}(x)} \right] &=& 0, \label{sp-sym-b} \\
\frac{d}{dx} \left[ \frac{dh_1^{\prime}}{dx}
 e^{2{\tilde \theta}(x)}\right] &=& 0, \label{sp-sym-c} \\
\frac{d}{dx} \left[ \frac{dh_1^{\prime\prime}}{dx}
 e^{2{\tilde \theta}(x)}\right] &=& 0. \label{sp-sym-d}
\end{eqnarray}
Here, $\kappa = 2iV_{\rm sp}/g$. Variation over $V$ yields the (first) self-consistency equation
\begin{eqnarray}
\label{sc-sym}
\pi \int dx e^{{\tilde \theta}(x)} = 2\lambda.
\end{eqnarray}
Let us first consider Eq. (\ref{sp-sym-c}). It is equivalent to
\begin{eqnarray}
\label{zero}
\frac{dh_1^{\prime}}{dx} = C\; e^{-2{\tilde \theta}(x)}.
\end{eqnarray}
Integrating it over $x\in (-1/2,+1/2)$, and taking into account the periodicity
of $h_1^{\prime}(x)$ we conclude that $C \equiv 0$, so that
\begin{eqnarray}
\label{h1p}
\frac{dh_1^{\prime}}{dx} = 0.
\end{eqnarray}
Analogously, we conclude from Eq. (\ref{sp-sym-d}) that 
\begin{eqnarray}
\label{h1pp}
\frac{dh_1^{\prime\prime}}{dx} = 0.
\end{eqnarray}

Equations (\ref{h1p}) and (\ref{h1pp}) enable us to write the instanton free
energy ${\tilde F}_v(\lambda) = F[Q_{\rm ins};V_{\rm sp},0] + iV_{\rm sp}\lambda$
obtained by minimizing Eq. (\ref{limit-sym}) in the form
\widetext
\Lrule
\begin{eqnarray}
{\tilde F}_v(\lambda) = \frac{\pi g}{2} \int dx \left\{
\left( \frac{d{\tilde \theta}}{dx}\right)^2 + \frac{1}{4}e^{2{\tilde \theta}(x)}
\left( \frac{dh}{dx}-1\right)^2 - \frac{\kappa}{2}e^{{\tilde \theta}(x)}
\right\} + \frac{1}{2}g\kappa\lambda,
\label{fe-sym-ap}
\end{eqnarray}
\Rrule
\narrowtext
\noindent
where the functions ${\tilde \theta}(x)$ and $h(x)$ obey Eq. (\ref{sp-sym-b}) and
the equation
\begin{eqnarray}
\frac{d^2{\tilde\theta}}{dx^2} + \frac{1}{4} e^{{\tilde \theta}(x)}
\left[
\kappa - \left( \frac{dh}{dx}-1 \right)^2 e^{{\tilde \theta}(x)}
\right] = 0
\label{sp-new}
\end{eqnarray}
that follows from Eqs. (\ref{sp-sym-a}), (\ref{h1p}) and (\ref{h1pp}).

Applying now Eq. (\ref{sp-new}) to Eq. (\ref{fe-sym-ap}), and taking into account the 
periodicity of ${\tilde \theta}^{\prime}(x)$ and the self-consistency condition Eq. (\ref{sc-sym}),
we come down to
\begin{eqnarray}
{\tilde F}_v(\lambda) = \frac{\pi g}{2} \int dx \left( \frac{d{\tilde\theta}}{dx} \right)^2
+ \frac{1}{4} g\kappa\lambda.
\label{fe-sym-final}
\end{eqnarray}
At this point we notice that Eqs. (\ref{fe-sym-final}), (\ref{sp-new}), (\ref{sp-sym-b}) 
and (\ref{sc-sym}) coincide with Eqs. (\ref{sp-u-eq4}), (\ref{sp-u-1eq}), (\ref{sp-u-2eq}) 
and (\ref{sp-u-eq3}), respectively. This proves that the universal, nonperturbative in $g^{-1}$,
corrections to the distribution of the level velocities at $\beta=4$ coincide 
with those found with the exponential accuracy for $\beta = 2$.

\widetext

\begin{references}
\bibitem{Rem1} The Gaussian distribution of level velocities, Eq. (\ref{vel-rmt}), is 
insensitive to the symmetry parameter $\beta$ only if the perturbation does not change the underlying symmetry 
of the unperturbed system.
\end{references}
\end{document}